# Estimating a Large Drive Time Matrix between Zip Codes in the United States: A Differential Sampling Approach[1]

Yujie Hu[1,2], Changzhen Wang[3], Ruiyang Li[4], Fahui Wang[3*]

[1]Department of Geography, University of Florida, Gainesville, FL 32611
[2]UF Informatics Institute, University of Florida, Gainesville, FL 32611
[3]Department of Geography & Anthropology, Louisiana State University, Baton Rouge, LA 70803
[4]Children's Environmental Health Initiative, Rice University, Houston, TX 77005

**Abstract.** Estimating a massive drive time matrix between locations is a practical but challenging task. The challenges include availability of reliable road network (including traffic) data, programming expertise, and access to high-performance computing resources. This research proposes a method for estimating a nationwide drive time matrix between ZIP code areas in the U.S.—a geographic unit at which many national datasets such as health information are compiled and distributed. The method (1) does not rely on intensive efforts in data preparation or access to advanced computing resources, (2) uses algorithms of varying complexity and computational time to estimate drive times of different trip lengths, and (3) accounts for both interzonal and intrazonal drive times. The core design samples ZIP code pairs with various intensities according to trip lengths and derives the drive times via Google Maps API, and the Google times are then used to adjust and improve some primitive estimates of drive times with low computational costs. The result provides a valuable resource for researchers.
**Keywords:** drive time matrix, ZIP code area, Google Maps API, intrazonal drive time, random sampling

## Introduction

Estimating a drive time matrix between locations is a critical task in spatial analysis, commonly encountered by researchers in geography, urban planning, transportation engineering, business management, and operational research, etc. To list a few, analytical models such as spatial interaction modeling (Simini et al., 2012), travel demand estimation (McFadden, 1974), location-allocation problems (ReVelle and Swain, 1970; Hu et al., 2019), spatial accessibility measures (Luo and Wang, 2003; Dony et al., 2015; Balomenos et al., 2019; Zhu et al., 2020), and delineation of health care market areas (Wang et al. 2020), rely on attainment of reliable drive time estimation from a set of origin locations to a set of destination locations. Such a task for a small-size matrix has become routine in many GIS and transportation packages, such as ESRI® ArcGIS and Caliper® TransCAD. However, it can be a challenging task for a large matrix.

---

[1] This is a preprint of: Hu, Y., Wang, C., Li, R., & Wang, F. (2020). Estimating a large drive time matrix between ZIP codes in the United States: A differential sampling approach. *Journal of Transport Geography*, 86, 102770. https://doi.org/10.1016/j.jtrangeo.2020.102770

Most studies in this scope focus on small areas like individual cities or counties. Estimating the time matrix at larger geographic scales, such as a national scope, is of increasingly great importance to researchers and policy makers. Take the U.S. health care as an example, a national travel time matrix is the most critical component to study the average geographic access to health care (Onega et al., 2008; Boscoe et al., 2012), measure the variations in geographic access across regions (Onega et al., 2017), identify the areas where the health care geographic access is significantly lower than average, suggest the most appropriate sites for new care facilities, and facilitate the implementation of other strategies for reducing health care disparities such as remote health care, health care on wheels, and transit to care.

The challenges for calibrating a large drive time matrix include availability of reliable road network (including traffic) data, programming expertise, and computational power. Many studies assume a free flow condition on roads to eliminate the data requirement on traffic, and are often limited to estimation of drive times from areas (e.g., ZIP code area, census tract) to the nearest locations (Onega et al., 2008; 2017; Boscoe et al., 2012; Ikram et al., 2015) or between locations within a short range (Shi et al., 2012) with a significantly reduced number of OD (origin-destination) pairs. Most recently, Saxon and Snow (2019) estimated a drive time matrix from each census tract to each primary care location within 62 miles (100 km) in the U.S. by tapping into advanced computing resources such as distributed computing and sophisticated algorithms, which may not be accessible by most researchers. They did not consider traffic conditions or node impedance, and the result tended to underestimate drive times especially for short trips.

One way to account for traffic effect in drive time estimation relies on the utilization of traffic sensors or auxiliary data sources. These sensors include loop detectors (Kwon et al., 2000; Coifman, 2002) and automatic vehicle identification systems—such as toll collection system (El Faouzi et al., 2009), license plate recognition system (van Hinsbergen et al., 2009), and Bluetooth-based system (Bhaskar and Chung, 2013)—that are installed at certain locations along the road. They can accurately capture travel speeds and times. Other sensors are rather flexible, such as the floating or probe vehicles that consist of a sample of vehicles equipped with GPS units running with traffic (De Fabritiis et al., 2008). Based on collected information on a vehicle's location, direction, and speed in a short time interval, drive times between any two locations in a network can be readily attained (Semanjski, 2015). A few recent studies attempted to estimate drive times using big data. Toole et al. (2015) used call detail records (CDRs) to obtain drive times for all road segments in five selected cities worldwide. Woodard et al. (2017) derived drive times on all roadways in the Seattle metropolitan region based on collected mobile phone GPS data. Although these methods can provide highly accurate estimates on actual drive times in traffic, their reliance on installation of physical equipment or big crowdsourced data restricts their usage to only small areas ranging from major corridors to metro regions.

An alternative approach is to use third-party web mapping services such as Google Maps and MapQuest. For example, Google Maps Distance Matrix API uses the Google data such as its road network and collected traffic information to estimate drive times between a set of origins and a set of destinations. Similarly, MapQuest's Route Matrix API uses open-source mapping data from the OpenStreetMap project to achieve this goal. Another benefit of using these commercial web services is to relieve analysts of the burden of preparing street network data and accessing GIS/transportation software (Boscoe et al., 2012). However, the free usage of these services comes with request limits. For instance, Google Maps offers free usage up to 40,000 OD records per month (Hu and Downs, 2019). A similar restriction applies to MapQuest. As a result, researchers usually use this approach to derive drive times for a limited number of OD pairs (Wang and Xu, 2011).

Another issue related to drive time estimation between areas is the so-called aggregation error (Hu and Wang, 2016). The centroid-to-centroid approach assumes that all people in an area live at the centroid of an area (Hewko et al., 2002), and inevitably overlooks intrazonal travels (Kordi et al., 2012; Bhatta and Larsen, 2011). For example, the average commuting time within a traffic analysis zone (TAZ) is 11.3 minutes for auto drivers in Cleveland, Ohio (Wang, 2003). Given the average area of 82.25 square miles for the ZIP code areas in the U.S., intrazonal drive times at the ZIP code area level can be significant, especially in low-density suburban or rural areas. Its omission accounts for a high percentage in error for short-range trips. A common approach approximates intrazonal travel distance as the radius of an area-equivalent circle (Frost et al., 1998, Horner and Murray, 2002; Hu and Wang, 2015). Some recent studies use Monte Carlo simulation techniques to improve the estimation of trip lengths between area units (Hu and Wang, 2016; Hu et al. 2017), and offers a viable solution to intrazonal drive time estimation.

This study seeks to estimate a very large drive time matrix between ZIP code areas in the U.S. ZIP code area is a popular geographic unit used in many nationwide datasets. For instance, ZIP code area is often the finest geographic scale at which health information is compiled and distributed in the U.S. (Berke and Shi, 2009). Our method (1) does not rely on intensive efforts in data preparation or access to advanced computing resources, (2) uses algorithms of varying complexity and computational time to estimate drive times of different trip lengths, and (3) accounts for both interzonal and intrazonal drive times. Both the program and results will be available for free download, and provide a valuable resource for researchers.

**Methodology**

Data used in this research are all publicly available. The GIS layers include road networks, ZIP Code Tabulation Areas (ZCTAs) (as a surrogate for ZIP code areas), and census blocks with the 2010 demographic data for the entire U.S. They are extracted from the TIGER Products from the U.S. Census Bureau (2019). The block layer with the 2010 population data is used to generate population weighted centroids for ZIP code areas, which are more accurate representation of their locations than their geographic centroids (Wang, 2015:78). There are 32,840 ZIP code areas in this study. All data processing is performed in ESRI® ArcGIS 10.6.

Figure 1 outlines the workflow for measuring the nationwide ZIP-to-ZIP drive time matrix. The process includes three major steps: (1) obtaining a preliminary estimate of the centroid-to-centroid drive times between every two ZIP code areas, (2) using Google Maps to derive drive times for randomly-sampled OD pairs and adjusting the full drive time matrix based on regression models, and (3) incorporating the intrazonal drive times associated with both origin and destination ZIP code areas to finalize the estimation. Each of these steps is described as an algorithm in detail below.

Three distance measures are used in the algorithms. *Geodesic distance* is the distance between two points through a great circle along the surface of a sphere that approximates the Earth's shape. It considers terrain curvature and thus is favored over the Euclidean distance as a primitive baseline estimate for long-distance drive time (Griffith et al., 2012; Wang, 2015: 28). The geodesic distance $d_{ij}$ can be formulated as the following (Wang, 2015: 28):

$$d_{ij} = \cos^{-1}(\sin b \cdot \sin d + \cos b \cdot \cos d \cdot \cos(c - a)) \cdot R$$

where $(a, b)$ and $(c, d)$ represent the pair of longitude and latitude in radians of location $i$ and $j$, respectively, and $R$ is the average radius of the Earth. *Network distance* is the distance through the shortest path via a road network. Finally, *network time* is the travel time (here drive time) through the fastest path via a road network.

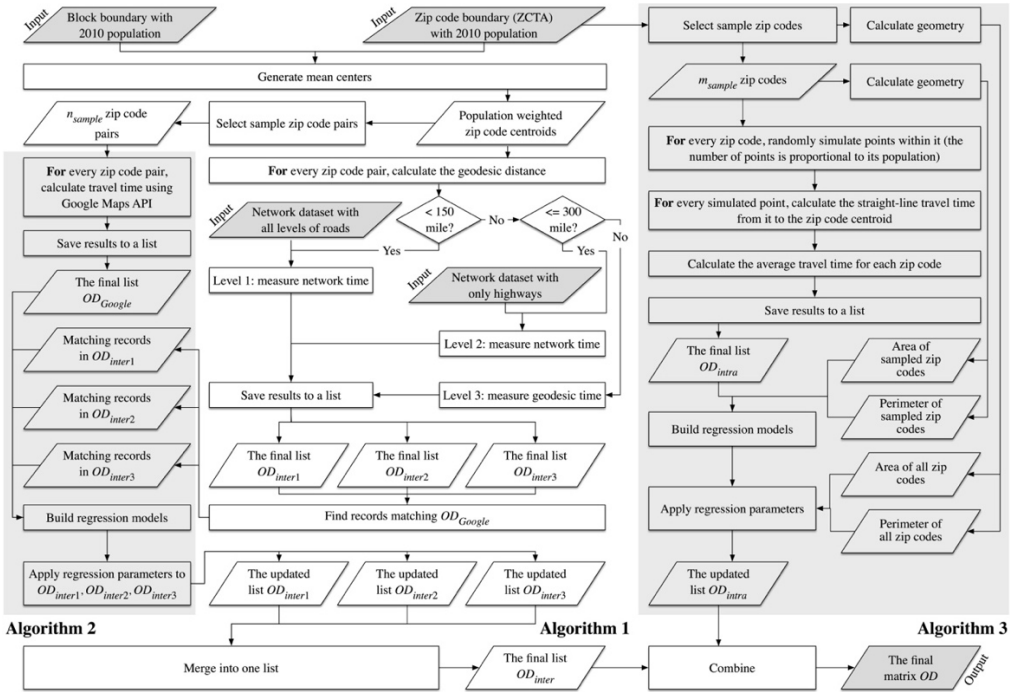

Figure 1. Workflow for estimating a ZIP-to-ZIP drive time matrix in the U.S.

*Algorithm 1: preliminary estimate of interzonal times*
Algorithm 1 is developed to provide a preliminary estimate of interzonal drive times for each OD pair of ZIP codes nationwide. It consists of three levels with incremental measurement accuracy corresponding to various trip lengths. This hierarchical design is to account for varying desirability of accuracy in drive time estimation and find a balance between our need for accuracy and fast computation. For short trips, travelers consider all levels of roads in routing since their spatial behaviors are sensitive to a minor difference in drive times. For medium-range trips, most parts are completed via major roads as travelers may not plan the drive times down to precise minutes. For very long trips, travelers may only need a ballpark number in drive times to assist their decisions. Therefore, this study designs three levels of trip lengths for our purpose. Level 1 is for short-range ZIP code pairs with the highest estimation accuracy, Level 2 for middle-range ZIP code pairs with moderate accuracy, and Level 3 for long-range pairs with the least accuracy. Considering a substantial number of ZIP code pairs being long-range at the national scale, this hierarchical design drastically reduces the computation time and enables us to calibrate the national ZIP-to-ZIP time matrix on a desktop PC within a reasonable amount of time.

Level 1 utilizes the complete road network including interstates, U.S. and state highways, major roads, and local roads to calculate drive times for short-range ZIP code pairs. For computational efficacy, ZIP code pairs are screened by geodesic distance to be assigned to Level 1 if they are within 150 miles (241 km) in geodesic distance, which is equivalent to 3 hours apart with a constant speed of 50 mph (80 km/h). The 3-hour drive time cutoff is often used by health care analysts as six hours are considered the limit for a patient to make a round trip between one's home and a health care facility and obtain services in one day (Shi et al. 2012; Onega et al. 2017). For each eligible OD pair, both trip ends are snapped onto the closest drivable streets within a 3.11-mile (5-km) search threshold. The network drive times between the snapped network locations are then measured via ArcGIS Network Analyst.

Level 2 measures the network drive times for medium-range ZIP code pairs by utilizing a road network of only interstates, U.S. and state highways. The screening for assignment of ZIP code pairs to this level is based on a range of 150-300 mile (241-483 km) in geodesic distance (equivalent to 3-6 hours with a 50 mph speed). Using the same search threshold, both origin and destination ZIP code centroids are snapped onto the simplified road network, and the network drive times are then estimated. As listed in Table 1, compared to Level 1, Level 2 trims the road network to a reasonable level of details and hence demands less computation time for each OD pair.

Table 1. Road network comparison between Level 1 and Level 2

|  | File size | Number of segments | Number of nodes |
|---|---|---|---|
| Level 1 network (all roads) | 29.3 GB | 18,368,450 | 64,138,323 |
| Level 2 network (simplified network) | 4.18 GB | 265,122 | 9,541,533 |

The remaining great number of long-range trips at Level 3 for ZIP code pairs of more than 300 miles (i.e., 6 hours) apart in geodesic distance. It is difficult for analysts with limited computational resources to estimate drive times for the OD pairs in this level via a road network—even one with reduced complexity. As stated previously, accuracy in drive times for these distant OD pairs is also less important. Therefore, this study uses the geodesic distance with a constant 50 mph speed to establish a baseline estimate of drive times in this level.

Integration of the results from the three levels yields a massive OD time matrix with 1,078,432,760 (= 32,840 * 32,839) records. A total of 32,840 (=32,840 * 1) records are missing due to the fact that most routing software packages, such as ArcGIS Network Analyst and Google Maps API, cannot measure travel distances or times from a place to itself. These remaining records are essentially intrazonal trips and will be estimated by Algorithm 3. Each record saves the origin ZIP code, destination ZIP code, preliminary estimated drive time (in minutes), distance (in miles), and the hierarchical level where time and distance are measured. Refer to Table 2 for more details, where ZIP code is simply referred to as "ZC" in the algorithms hereafter.

Table 2. Algorithm 1 for a preliminary measure of interzonal drive times

**Input**:
1) A point layer of national ZIP code centroids (population weighted) $ZC = \{zc_1, zc_2, ..., zc_m\}$, where $m \in [1, 32,840]$; each centroid $zc_m$ contains the ID and 2010 population of that ZIP code area;
2) A network dataset of national street centerlines (of various levels) $S = \{SL_1, SL_2\}$, where $SL_1$ is a set that includes interstates, interstate toll roads, freeway ramps, and state highways, and $SL_2$ contains major roads and local streets;

**Output**:
A massive OD cost matrix $OD = \{od_1, od_2, ..., od_n\}$, where $n \in [1, 1,078,432,760]$; each OD pair is a set containing $\{zc_i, zc_j, t_{ij}, d_{ij}, hl\}$, where $i \in [1, 32,840]$, $j \in [1, 32,840]$ and $j \neq i$, $t_{ij}$ and $d_{ij}$ represent the estimated drive time and distance, respectively, and $hl$ denotes the hierarchical level in which $t_{ij}$ and $d_{ij}$ are calculated;

**Variables**:
1) $d_{max}$: the maximum search distance in miles to locate (snap) a ZIP code centroid $zc_i$ onto the road network $S$; it is set to 3.11 miles;
2) $Os\_hl_1$: a temporary set of origin ZIP code centroids that are located onto $S$ for hierarchical Level 1;
3) $Ds\_hl_1$: a temporary set of destination ZIP code centroids that are located onto $S$ for hierarchical Level 1;
4) $Os\_hl_2$: a temporary set of origin ZIP code centroids that are located onto $S$ for hierarchical Level 2;
5) $Ds\_hl_2$: a temporary set of destination ZIP code centroids that are located onto $S$ for hierarchical Level 2;
6) $Os\_hl_3$: a temporary set of origin ZIP code centroids that are located onto $S$ for hierarchical Level 3;
7) $Ds\_hl_3$: a temporary set of destination ZIP code centroids that are located onto $S$ for hierarchical Level 3;
8) $OD\_notfound$: a list $\{zc_i, zc_j, t_{ij}, d_{ij}, hl\}$ containing OD pairs whose drive times and distances cannot be resolved;
9) $d\_hl_1$: the geodesic distance threshold in mile for hierarchical Level 1; it is set to 150;
10) $d\_hl_2$: the geodesic distance threshold in mile for hierarchical Level 2; it is set to 300;
11) $ds$: a constant driving speed; it is set to 50 mph (or 0.83 mile per minute);

**Functions**:
1) **network_dist**($zc_i$, $zc_j$, $S$): measure network distance through road network $S$;
2) **network_time**($zc_i$, $zc_j$, $S$): measure network drive time through $S$;
3) **geodesic_dist**($zc_i$, $zc_j$): calculate geodesic distance;
4) **geodesic_time**($zc_i$, $zc_j$): calculate geodesic drive time, which is the ratio of the estimated geodesic distance and $ds$;

**Steps**:
1) Initialize $Os\_hl_1$, $Os\_hl_2$, $Os\_hl_3$, $Ds\_hl_1$, $Ds\_hl_2$, $Ds\_hl_3$, and $OD\_notfound$;
2) **for** a ZIP code centroid $zc_i$ in $ZC$, $i \in [1, 32,840]$:
    Empty $Os\_hl_1$, $Os\_hl_2$, $Os\_hl_3$, $Ds\_hl_1$, $Ds\_hl_2$, and $Ds\_hl_3$;
    **for** a ZIP code centroid $zc_j$ in $ZC$, $j \in [1, 32,840]$ and $j \neq i$:
        **if geodesic_dist**($zc_i$, $zc_j$) < $d\_hl_1$:
            Assign $S = \{SL_1, SL_2\}$;
            Snap $zc_i$ onto $S$ within $d_{max}$ and add $zc_i$ into $Os\_hl_1$;
            Snap $zc_j$ onto $S$ within $d_{max}$ and add $zc_j$ into $Ds\_hl_1$;
            Measure $t_{ij}$ = **network_time**($zc_i$, $zc_j$, $S$) and $d_{ij}$ = **network_dist**($zc_i$, $zc_j$, $S$) between $zc_i$ in $Os\_hl_1$ and $zc_j$ in $Ds\_hl_1$;
            **if** $t_{ij}$ is NULL or $d_{ij}$ is NULL:
                Add the list $\{zc_i, zc_j, t_{ij}$ = NULL, $d_{ij}$ = NULL, $hl = 1\}$ to $OD\_notfound$;
            **else**:
                Add the list $\{zc_i, zc_j, t_{ij}, d_{ij}, hl = 1\}$ to the final OD cost matrix $OD$;
        **else if** $d\_hl_1 \leq$ **geodesic_dist**($zc_i$, $zc_j$) $\leq d\_hl_2$:
            Assign $S = \{SL_1\}$;
            Snap $zc_i$ onto $S$ within $d_{max}$ and add $zc_i$ into $Os\_hl_2$;
            Snap $zc_j$ onto $S$ within $d_{max}$ and add $zc_j$ into $Ds\_hl_2$;
            Measure $t_{ij}$ = **network_time**($zc_i$, $zc_j$, $S$) and $d_{ij}$ = **network_dist**($zc_i$, $zc_j$, $S$) between $zc_i$ in $Os\_hl_2$ and $zc_j$ in $Ds\_hl_2$;
            **if** $t_{ij}$ is NULL or $d_{ij}$ is NULL:
                Add the list $\{zc_i, zc_j, t_{ij}$ = NULL, $d_{ij}$ = NULL, $hl = 2\}$ to $OD\_notfound$;
            **else**:
                Add the list $\{zc_i, zc_j, t_{ij}, d_{ij}, hl = 2\}$ to $OD$;
        **else**:
            Add $zc_i$ into $Os\_hl_3$;
            Add $zc_j$ into $Ds\_hl_3$;
            Measure $t_{ij}$ = **geodesic_time**($zc_i$, $zc_j$) and $d_{ij}$ = **geodesic_dist**($zc_i$, $zc_j$) between $zc_i$ in $Os\_hl_3$ and $zc_j$ in $Ds\_hl_3$;
            Add the list $\{zc_i, zc_j, t_{ij}, d_{ij}, hl = 3\}$ to $OD$;
3) Empty $Os\_hl_1$, $Os\_hl_2$, $Os\_hl_3$, $Ds\_hl_1$, $Ds\_hl_2$, and $Ds\_hl_3$;
4) **for** lists $\{zc_i, zc_j, t_{ij}$ = NULL, $d_{ij}$ = NULL, $hl\}$ in $OD\_notfound$:
    **if** $hl = 1$:
        Assign $S = \{SL_1\}$;

Snap $zc_i$ onto $S$ within $d_{max}$ and add $zc_i$ into $Os\_hl_2$;
                    Snap $zc_j$ onto $S$ within $d_{max}$ and add $zc_j$ into $Ds\_hl_2$;
                    Measure $t_{ij}$ = **network_time**($zc_i$, $zc_j$, $S$) and $d_{ij}$ =
                    **network_dist**($zc_i$, $zc_j$, $S$) between $zc_i$ in $Os\_hl_2$ and $zc_j$ in $Ds\_hl_2$;
                    **if** $t_{ij}$ is NULL or $d_{ij}$ is NULL:
                        Measure $t_{ij}$ = **geodesic_time**($zc_i$, $zc_j$) and $d_{ij}$ =
                        **geodesic_dist**($zc_i$, $zc_j$);
                        Add the list {$zc_i$, $zc_j$, $t_{ij}$, $d_{ij}$, $hl = 3$} to $OD$;
                    **else**:
                        Add the list {$zc_i$, $zc_j$, $t_{ij}$, $d_{ij}$, $hl = 2$} to $OD$;
                **if** $hl = 2$:
                    Measure $t_{ij}$ = **geodesic_time**($zc_i$, $zc_j$) and $d_{ij}$ = **geodesic_dist**($zc_i$, $zc_j$);
                    Add the list {$zc_i$, $zc_j$, $t_{ij}$, $d_{ij}$, $hl = 3$} to $OD$;
    5) Write $OD$ to a file.

*Algorithm 2: calibrating interzonal times on randomly sampled OD pairs by Google Maps API*

In essence, Algorithm 1 returns free-flow drive times without considering traffic and road congestion for short- and medium-range OD pairs, and yields a very primitive baseline estimate for long-range OD pairs. The time estimates tend to be downward biased. Algorithm 2 improves the estimates by using Google Maps API to account for actual experiences on the road including traffic condition (Wang and Xu, 2011). Google's most recent pay-as-you-go pricing plan supports free usage of Distance Matrix API up to 40,000 OD records per month (Hu and Downs, 2019). It is cost prohibitive to use this method to calibrate drive time for all OD pairs. We apply it only to a small subset of randomly-sampled OD pairs.

Algorithm 2 is described as the following:
1) randomly selecting a ZIP code centroid $zc_i$ from $ZC$ ($i \in [1, 32,840]$) as an origin location;
2) randomly choosing a ZIP code centroid $zc_j$ from $ZC$ ($j \in [1, 32,840]$ and $j \neq i$) as a destination location;
3) measuring drive time $t_{ij}$ and distance $d_{ij}$ from $zc_i$ to $zc_j$ by sending a request to Google Maps Distance Matrix API; and
4) adding the resulting list {$zc_i$, $zc_j$, $t_{ij}$, $d_{ij}$} into a file and going back to step 1) until the predefined sample size $n_{sample}$ is reached.

Once collected, the relations between drive times derived by Algorithms 1 and 2 on the sampled subset are established through three regression models, corresponding to the three hierarchical levels in Algorithm 1. The three empirically-derived regression models are then applied to the remaining OD pairs for adjusting the preliminary estimates of drive times (and distances). Both algorithms use Google Maps drive time as the true reference values for adjusting the estimation. When no departure time is specified, such time is average time across various traffic conditions independent of time of a day or day of a week.

*Algorithm 3: measuring intrazonal times*

Intrazonal trips (from a ZIP code area to itself) are usually neglected in time and distance estimation models (Kordi et al., 2012; Bhatta and Larsen, 2011), including Algorithms 1 and 2, because drive time and distance are often approximated through a centroid-to-centroid approach. Built upon the method developed by Hu and Wang (2016), Algorithm 3 is proposed to measure intrazonal drive times. Similar to Algorithm 2, this process is applied to a randomly-sampled subset of ZIP code areas, as described below:

1) randomly picking $m_{sample}$ ZIP code centroids and saving the corresponding ZIP code areas into a polygon layer $ZP = \{zp_1, zp_2, ..., zp_k\}$, where $k \in [1, m_{sample}]$; each polygon $zp_k$ contains the ID, 2010 population, area, and perimeter of that ZIP code zone;
2) for each selected ZIP code polygon $zp_i$ ($i \in [1, m_{sample}]$), generating random points in it using the Monte Carlo simulation method (Hu and Wang, 2016), and the number of random points is proportional to the population size of $zp_i$;
3) computing the distance from each random point to the ZIP code centroid $zc_i$ and then calculating the average distance $d_{ii}$ for $zc_i$;
4) measuring the average drive time $t_{ii}$ for $zc_i$ by dividing $d_{ii}$ by a constant travel speed of 25 mph (40 km/h), which reflects a slower speed on local roads within a ZIP code area; and
5) adding the resulting list $\{zc_i, t_{ii}, d_{ii}\}$ into a file and continuing for the next ZIP code until all selected ZIP codes are explored.

As intrazonal drive time and distance are reported to be positively related to the perimeter and area of a zone (Frost et al., 1998; Horner and Murray, 2012; Hu and Wang, 2019), several regression models are tested to identify the relations. The best fitting model is then applied to populate intrazonal drive times and distances for the remaining ZIP code areas.

Finally, the intrazonal times (and distances) are integrated into the interzonal estimates for updating the entire OD time matrix. The final drive time $t_{ij}'$ between each OD pair $zc_i$ and $zc_j$ includes three components—intrazonal time $t_{ii}$ in $zc_i$, interzonal time $t_{ij}$, and intrazonal time $t_{jj}$ in $zc_j$. Specifically, it is defined as: $t_{ij}' = t_{ii} + t_{ij} + t_{jj}$. See Figure 2 for an illustration where shaded areas represent zip code zones and points within a shaded area denote Monte Carlo simulated individuals within that zip code zone.

The above three algorithms are coded in Python. All the analyses are carried out in a desktop PC with an Intel Core i7 processer and a 16 GB RAM.

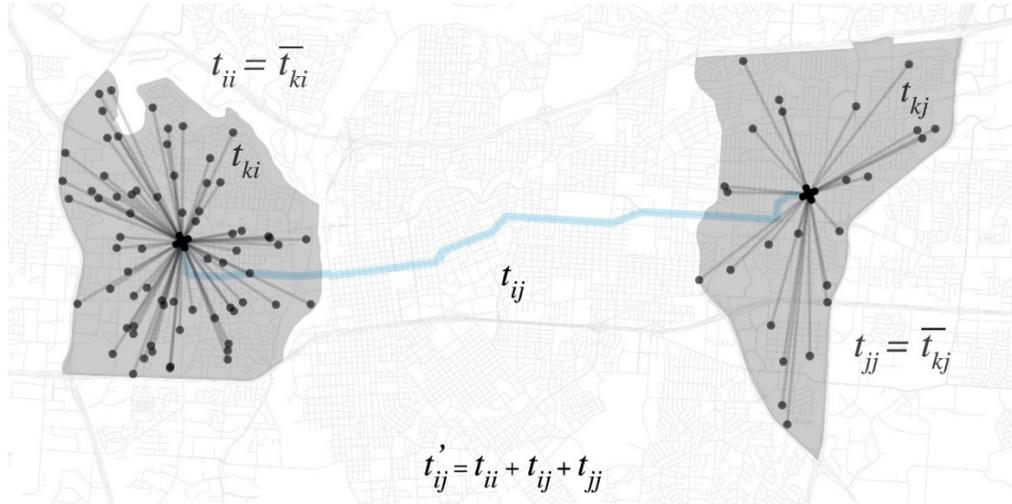

Figure 2. Illustration of the drive time components

**Results**

Based on the 2010 block population data, population weighted centroids for the 32,840 ZIP codes in the U.S are created. Together with the two network datasets of different road levels, they are fed into Algorithm 1. This process results in a massive OD cost matrix of 1,078,432,760 records, each of which consists of estimation of drive time and distance for a ZIP code pair. It takes about 76 hours for Algorithm 1 to compute and export the national ZIP-to-ZIP time matrix, and the breakdowns are: Level 1 consumes 6 hours, Level 2 60 hours, and Level 3 10 hours.

Algorithm 2 is implemented on a randomly-sampled subset of OD pairs derived from Algorithm 1. Given the free usage constraints discussed previously, Algorithm 2 is run to collect Google drive times over a period of four months. It finally gathers 124,350 ($n_{sample}$) valid ZIP-to-ZIP trips, which are associated with 32,478 unique ZIP codes (about 99% of the national 32,840 ZIP codes). Each trip includes Google's estimates of drive time and distance. Table 3 lists the frequency distribution of OD trips derived from Algorithms 1 and 2. The sampling intensity for the short-range trips (Level 1) is about triple those medium-range trips (Level 2) and quadruple those long-range trips (Level 3). Oversampling shorter trips is to enhance their representation and ensure the quality of subsequent interpolation for greater interests in acquiring shorter-range travel times. Figure 3 shows sampled ZIP code pairs of Level 1.

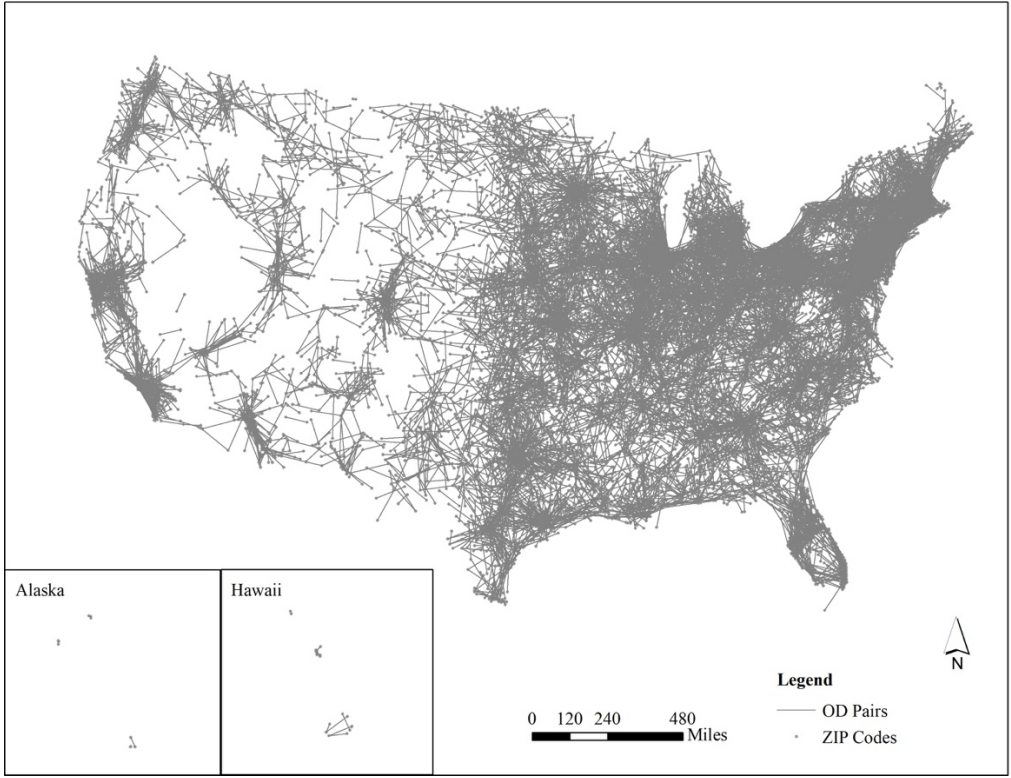

Figure 3. Sampled ZIP code pairs of Level 1

Table 3. Frequency distribution of OD pairs derived from Algorithms 1 and 2

| Level | Algorithm 1 | | Algorithm 2 | | Sampling % |
| --- | --- | --- | --- | --- | --- |
| | Count ($n$) | % | Count ($n_{sample}$) | % | ($n_{sample}/n$) |
| 1 | 28,241,488 | 2.62 | 11,684 | 9.40 | 0.041 |
| 2 | 80,547,438 | 7.47 | 11,365 | 9.14 | 0.014 |
| 3 | 969,643,834 | 89.91 | 101,301 | 81.46 | 0.010 |
| Total | 1,078,432,760 | 100.00 | 124,350 | 100.00 | 0.012 |

Figures 4A, 4C, and 4E (and 4B, 4D, 4F) plot drive times (distances) estimated by Algorithm 1 against times (distances) by Algorithm 2 at Level 1, 2 and 3, respectively. Only the data points within 24 hours (1,200 miles, or 1,931 km) for Level 3 are shown. Clearly, the pattern is largely consistent between the two algorithms for ZIP code pairs across the three levels. A few observations merit discussion. It is obvious in Figures 4A and 4C that Algorithm 1 tends to underestimate travel times. One likely cause for this trend is the omission of congestion in Algorithm 1. Interestingly, such a trend of downward estimates is also observed in distance measurements in Figures 4B and 4F. This is because the distances returned by Algorithm 2 are simply the lengths in mileage of those quickest routes in terms of drive times, which are commonly through highways. Therefore, distances measured by Algorithm 1—the "shortest" routes in terms of mileage—are consistently lower than the values estimated by Algorithm 2. The exact fitting power of each needs to be examined by regression.

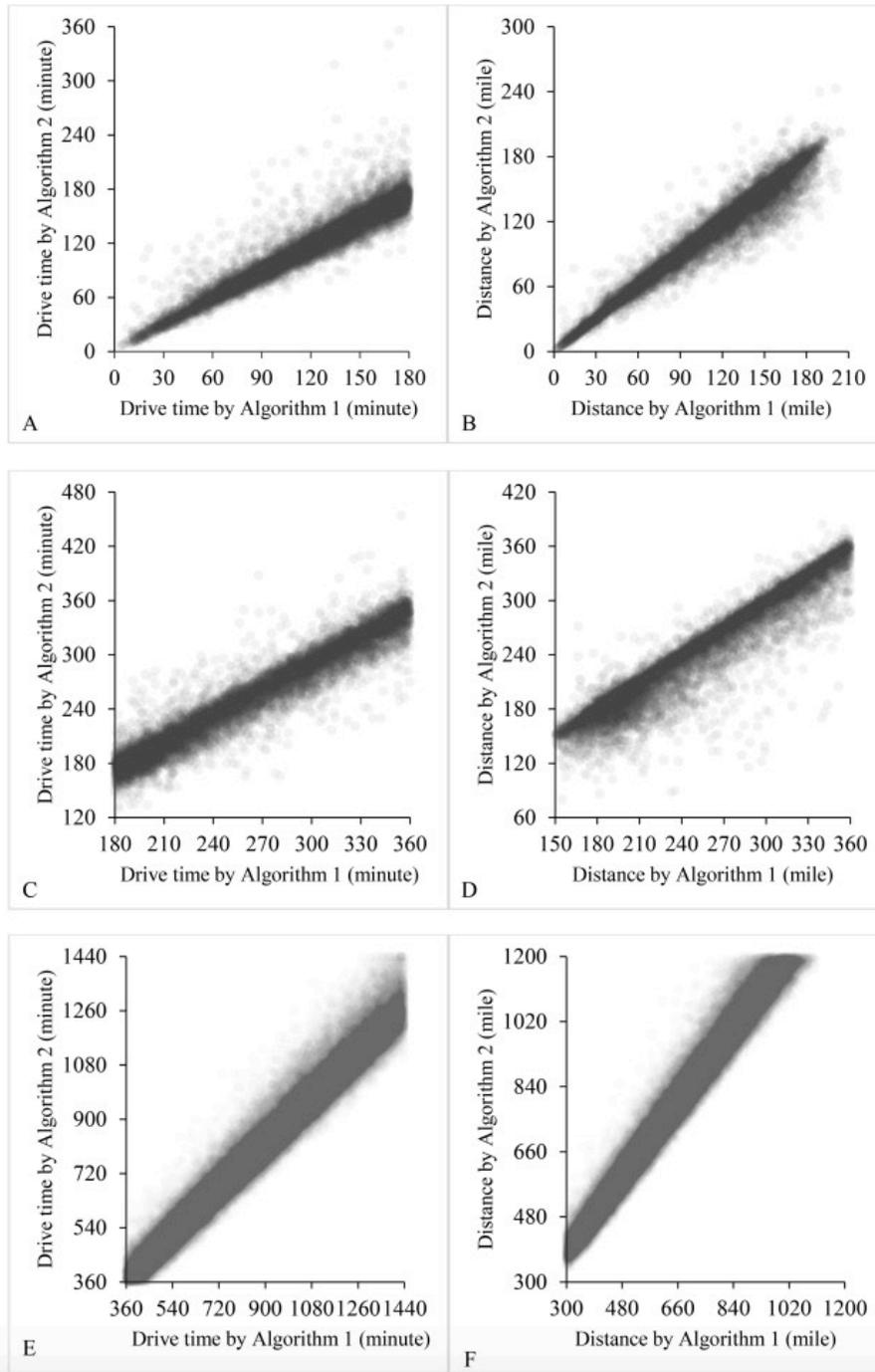

Figure 4. A, C and E represent drive times by Algorithm 1 vs. Algorithm 2 at Level 1, 2 and 3, respectively; B, D and F represent travel distances by Algorithm 1 vs. Algorithm 2 at Level 1, 2 and 3, respectively

Corresponding to each of the three levels in Algorithm 1, a regression is run to infer the relationship between Google drive times by Algorithm 2 (dependent variable) and Algorithm 1 drive times (independent variable). Results are summarized in Table 4. It shows that Algorithm 1 times explain the variation in the matched Google times by 91 percent at Level 1, 93 percent at Level 2, and 96 percent at Level 3. Table 4 also reports the results for regression models on distance measures from the two algorithms, with the $R^2 = 0.95$, 0.93, and 0.99 for Levels 1, 2 and 3, respectively. The higher fitting powers by the travel distance models than the drive time ones are due to the uncertainty in traffic congestion effect on drive time measurement. As travel distances are not used as often as drive times in measuring spatial impedance, our discussion focused on drive time.

Table 4. Regression models of drive time and distance estimated by Algorithms 1 and 2

|  | Algorithm 2 Drive time | | | Algorithm 2 Travel distance | | |
|---|---|---|---|---|---|---|
|  | Level 1 | Level 2 | Level 3 | Level 1 | Level 2 | Level 3 |
| Intercept | 5.69*** (16.65) | 0.32 (0.469) | 33.91*** (42.54) | 3.92*** (16.15) | -6.69*** (-9.72) | 20.86*** (47.26) |
| $T_{a1}$ | 0.94*** (347.689) | 0.96*** (375.45) | 0.88*** (1590.26) | 0.95*** (467.88) | 0.99*** (384.83) | 1.18*** (3214.65) |
| N | 11,684 | 11,365 | 101,301 | 11,684 | 11,365 | 101,301 |
| $R^2$ | 0.91 | 0.93 | 0.96 | 0.95 | 0.93 | 0.99 |

Note: $T_{a1}$ represents Algorithm 1 derived drive times and distances, and N denotes the number of observations.
***Significant at the 0.001 level.

The overall high $R^2$ values in the short-, medium-, and long-range models demonstrate the promise of Algorithm 1 for providing a fast and reliable estimate of ZIP-to-ZIP drive times nationwide. Why does the explanatory power increase from Level 1 to Level 2 and then to Level 3 as the predictor becomes coarser (i.e., from estimated drive times based on detailed road network, to simplified road network, to simply geodesic distance)? As stated previously, when the trip length increases, a larger error in absolute value does not necessarily correspond to a larger difference in relative value (e.g., percentage). Moreover, Google times may be more sensitive to the effect of traffic condition in shorter range trips. If geodesic distances are used as the explanatory variable for Google times in the corresponding samples in Level 1 and Level 2, the $R^2$ values are as low as 0.78 and 0.74, respectively. Similarly, a slightly lower $R^2$ (0.9115, to be precise) is observed if the predictor is drive times based on the simplified road network (with only major highways) than reported in Table 4 for the OD pair samples in Level 1 (0.9118, to be precise). To recap, the best predictor is estimated drive time based on the all-road network, followed by the only-highway network, and then simply geodesic distance. However, the gain in accuracy diminishes as distance increases. Therefore, it is computationally prudent to reduce the road network data requirement and computational complexity as we move from short-, to medium- and long-range OD distance pairs. It largely validates our strategy as currently designed.

The regression results are then used to adjust the preliminary estimates of times and distances on the rest ZIP code pairs in the national cost matrix derived by Algorithm 1. The final task is to estimate intrazonal drive times. Algorithm 3 is run over a random sample of 320 ($m_{sample}$) ZIP code areas (roughly 1% of the ZIP codes nationwide). For each selected ZIP code, the number of individuals (points) is simulated to equal its population in order to obtain more accurate estimates of mean intrazonal drive time in more populous ZIP code areas. Trips between simulated points are subsequently simulated in order to derive an average intrazonal drive time. As shown in Figures 5 and 6, the perimeter of a ZIP code is linearly correlated with its intrazonal time (or distance), whereas the area of a ZIP code has a nonlinear relationship with its intrazonal time (or distance). Therefore, as reported in Table 5, both a ZIP code's perimeter and square root of its area are significant explanatory variables in predicting intrazonal drive times. This model is then used to estimate intrazonal drive times for the remaining 32,520 (= 32,840 - 320) ZIP codes in the nation. The final drive time estimate $t'_{ij}$ of each ZIP code pair in the national time matrix is derived based on $t'_{ij} = t_{ii} + t_{ij} + t_{jj}$. In a similar way, the final travel distance matrix is obtained.

Table 5. Regression models on intrazonal drive time and distance

|  | Intrazonal drive time | Intrazonal travel distance |
|---|---|---|
| Intercept | -0.02 (-0.17) | -0.01 (-0.17) |
| $ZC_P$ | 0.04*** (6.61) | 0.02*** (6.62) |
| $\sqrt{ZC_A}$ | 0.88*** (18.39) | 0.37*** (18.39) |
| N | 320 | 320 |
| $R^2$ | 0.96 | 0.96 |

Note: $ZC_P$ represents perimeter of a ZIP code, $\sqrt{ZC_A}$ means the square root of a ZIP code's area, and N is the number of observations; t values in parentheses, ***Significant at the 0.001 level.

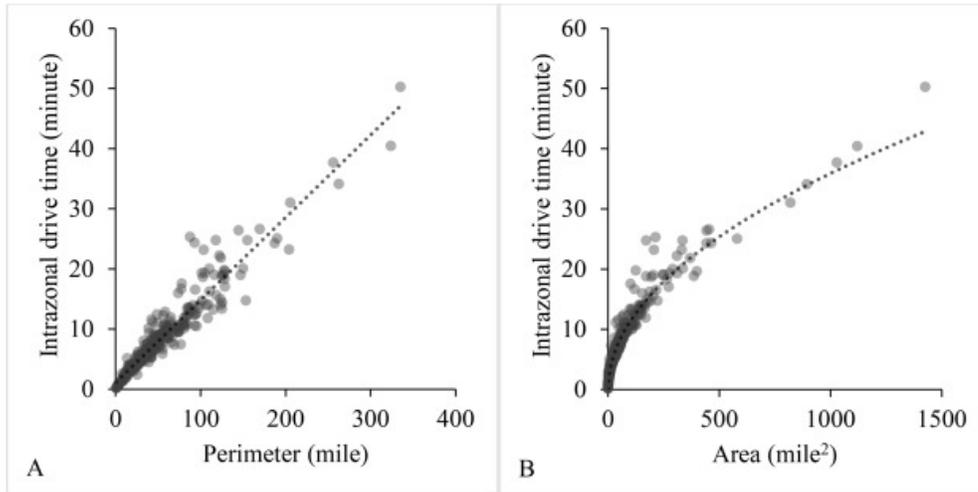

Figure 5. (A) Intrazonal drive time vs. perimeter, (B) intrazonal drive time vs. area

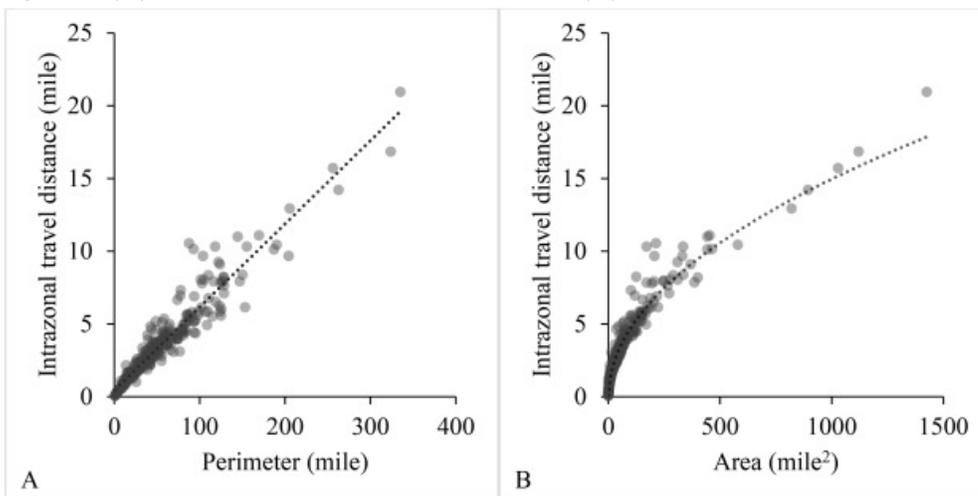

Figure 6. (A) Intrazonal travel distance vs. perimeter, (B) intrazonal travel distance vs. area

Originally designed by the United States Postal Service (USPS), ZIP code is a five-digit number corresponding to address points in the U.S., and such data have never been released by the USPS for public access. It is, therefore, not an areal unit where a ZIP code has a specified physical boundary. For the sake of spatial analysis, many entities have attempted to delineate ZIP code areas, such as the ZCTAs by the U.S. Census (Khan et al., 2019). Our recent experience in using this drive time matrix indicates that some USPS ZIP codes are not included in the ZCTAs data. Most of those missing ZIP codes have no associated area such as post office box ZIP codes and single site ZIP codes (government, building, or large volume customer). One may use the following algorithm to interpolate the drive times on corresponding missing OD pairs:

1) For drive time between a missing ZIP code (say, origin) and a known ZIP code (say, destination): identify the 3 nearest known ZIP codes from the missing ZIP code, locate the 3 drive times between each of the 3 nearby known ZIP codes (origin) and the known ZIP (destination) from the provided matrix, and use their average drive time as the one between the missing ZIP code and the known ZIP code.
2) For drive time between two missing ZIP codes: identify the 3 nearest known ZIP codes from the origin ZIP code and also the 3 nearest known ZIP codes from the destination ZIP code, locate the corresponding 9 drive times between each of the 3 nearby origins and each of the 3 nearby destinations from the provided OD cost matrix, and use their average drive time as the one between the two missing ZIP codes.

The above proposition assumes that the location of a missing ZIP code can be approximated by the average of its three nearest ZIP codes.

**Concluding comments**

This paper illustrates the estimation of the nationwide ZIP-to-ZIP drive time matrix in the U.S. The drive times derived by the Google Maps API on randomly-sampled OD pairs serve two purposes: facilitating empirical models to further improve the preliminary estimates based on road networks or simply geodesic distances, and validating our design of the methods of varying computational complexity and differential sampling intensity. As trip lengths increase, the approach requires less data preparation and uses less computational power without much compromising the quality of results.

Our own motivation for undertaking this endeavor is to facilitate a study that examines a national health care market structure. We hope that the derived matrix becomes an important resource for researchers who may need it in spatial analysis of a national scope or a large region. For instance, a recent study on measuring and improving accessibility to public libraries in the U.S. (Donnelly, 2015) could benefit from a more accurate measure of drive time from us. In addition, the estimated coefficients and other parameters from the regression models in both Algorithm 2 and Algorithm 3 can be used as a reference in other studies when such information at the national scale is not available. For studies being performed in other geographic scales, such as census tract, or other geographic areas, the derived parameters can be also referenced as a baseline. The proposed research method (or framework) is also useful for one to imitate in a different country (region) of a similar scale. The method has been wrapped into a convenient ArcGIS tool with a user interface, where researchers can easily select input data and make changes to key parameters, such as the constant travel speed, the predefined three hierarchical levels, and the number of requests sent to Google Maps API, to make the tool work for their own data. We will provide both the tool and the matrix for free download.

Several limitations of this study merit discussion. First, this research considers driving as the only transportation mode. The omission of other modes could be problematic especially for studies focusing on other trip purposes or in other areas where public transit service coverage is high. For example, the General Transit Feed Specification (GTFS) data can be integrated into the road network for calculating drive time by transit. In addition, potential users of the derived matrices are suggested to proceed with caution when using travel times of some medium- or long-range trips, such as from Alaska to the contiguous U.S., if they favor more accurate estimates down to minutes. Some of these trips are likely to be made by other modes such as air or train, which are not accounted for by the proposed approach. Another related issue is that the use of ferry is permitted in Algorithm 2 by default, which yields much shorter travel between areas separated by water, e.g., between Michigan and Wisconsin, or with island barriers in coastal areas, than otherwise. If it is desirable to avoid the use of ferry, one can simply specify one parameter (avoid="ferries") in Algorithm 2, according to the Google Maps API. In any case, the estimated drive times are a good proxy for travel impedance.

Secondly, more work is needed to improve the baseline estimation on the current division of three hierarchical levels in Algorithm 1. Instead of using people's perceptions, one may design a simulation procedure that examines the national road network and identifies at what distances it would be most appropriate to simplify the road network. Other types of times warrant consideration for more reasonable estimates, especially for long-range trips in Level 3, such as stopping time for bathroom breaks, gas refill, or sleep. Another issue may arise from the current selection of travel speeds, such as 50 mph in Algorithm 1 and 25 mph in Algorithm 3, in estimating drive times. These values may overestimate drive times of distant zip code pairs in Algorithm 1 or large zip code zones in Algorithm 3 in which case highways and interstates with higher speed limits are more likely to be utilized. Similarly, the selection of an appropriate distance threshold to snap locations onto a road network may depend heavily on the geography being studied. More experiments are needed to determine the most appropriate values. In addition, as discussed in Shi (2007), the Monte Carlo randomization in Algorithm 3 would benefit from a process that considers population distribution, such as the block-level population data, rather than the zip code zone itself. Such a finer geographic resolution would demand additional computation, however. Another potential improvement to Algorithm 3 could be the consideration of the number of road segments or the total length of road segments within ZIP code areas besides perimeter and area.

In addition, it would be worthwhile to report uncertainty along with the matrices. Three sources of uncertainty are relevant in this study: (1) the three defined hierarchical levels, (2) the centroid-based representation of a zip code zone, and (3) the random sampling of zip code pairs in Algorithm 2. For example, a possible solution to address the random sampling issue might be to consider different geographies and population sizes (Delmelle et al., 2019). Furthermore, other road network data sources such as the OpenStreetMap (OSM) could be employed, especially for regions or countries that do not have access to high-quality road network data. Finally, it is worthwhile to make the proposed method available in a non-ArcGIS environment since ArcGIS is not free to the public, especially for researchers in other counties.

**Acknowledgement**

Financial support from the National Cancer Institute (NCI), National Institutes of Health, under Grant R21CA212687, is gratefully acknowledged. Points of view or opinions in this article are those of the authors, and do not necessarily represent the official position or policies of NCI. Hu also would like to acknowledge the support by the Ralph E. Powe Junior Faculty Enhancement Awards from the ORAU (Oak Ridge Associated Universities). Comments from 3 anonymous reviewers helped us prepare a much improved final version of the paper.